# William Kruskal: My Scholarly and Scientific Model

**Stephen E. Fienberg**

When I arrived at the University of Chicago as an assistant professor in the summer of 1968, Bill Kruskal was department chair and he became a constant presence in my life, introducing me to new topics and people, gently advising me, encouraging me to look more deeply into almost everything we talked about. Many of the activities of my subsequent career, in statistics proper and at the interface with other fields, had their roots in my interactions with Bill during my time at Chicago.

My arrival occurred just before the Democratic convention to pick a candidate for that year's presidential elections. Over lunch one day I expressed to Bill an interest in the accuracy of public opinion polls and their scientific foundation. The next thing I knew Bill had recommended me to the producers of a university television interview program that was about to air on a local station. A group of faculty ended up doing three successive panel discussion programs on polling. Norman Bradburn and Ken Prewitt were part of this effort and I've continued to interact with both of them throughout my career. I also began to look carefully at the regular newspaper reports of the Chicago Sun–Times Poll, and Bill encouraged me to make a plan to assess its accuracy—this meant assembling a data set of predictions and of course election results. Before too long this became a working manuscript and Bill encouraged me to submit it to the *Journal of the American Statistical Association* (*JASA*) for publication. These activities grew into my later research on sample surveys.

During another lunch hour that first fall, Bill introduced me to Hans Zeisel at the Quadrangle Club and, within the week, Hans solicited my assistance analyzing data on the composition of the jury pool for the trial of Dr. Spock and others, which ended up first as a law journal article Hans wrote and then as part of a chapter in *Statistics: A Guide to the Unknown*, an American Statistical Association–National Council of Teachers of Mathematics (ASA–NCTM) collaborative volume for which Bill was one of the editors. I later used this example for a related ASA–NCTM project organized by Fred Mosteller, *Statistics by Example*. I also became a regular at Hans' quantitative methods seminar at the law school. It was here that I met Michael Finkelstein (a guest speaker), Norval Morris, Frank Zimring and others and was introduced to the study of criminal justice statistics and the fascinating interface between statistics and the law.

The Vietnam War was a major topic of conversation around the department and at faculty gatherings. Bill was fascinated by the regular data being shared on reported deaths of American soldiers and thought that there must be an interesting set of statistical issues there. When the draft lottery drawing took place in 1969, and a number of others claimed to find flaws in the "randomness" of the outcome, it was Bill who encouraged me to do some careful data analysis and to begin to develop a scholarly article that included the history of lotteries and the role of randomization. At first blush this didn't look like a logical piece for *The Annals of Mathematical Statistics* or for *JASA*, but Bill suggested that this would make a good article for *Science*, to which he had introduced me shortly after my arrival at Chicago and which he clearly read from cover to cover. This piece went through repeated revisions, with constant edits from Bill, and references to things I should explore, both in the analysis and in the scholarly treatment of the history. Long after my draft lottery article

*Stephen E. Fienberg is Maurice Falk University Professor of Statistics and Social Science in the Department of Statistics, the Center for Automated Learning and Discovery, and Cylab, all at Carnegie Mellon University, Pittsburgh, Pennsylvania 15213-3890, USA e-mail: fienberg@stat.cmu.edu.*

(Fienberg, 1971) was published and I was at Minnesota, and even later at Carnegie Mellon, I would get newspaper clippings from Bill on related topics.

One of the earliest journal submissions from my Ph.D. thesis, on the geometry of the $2 \times 2$ contingency table, was rejected by *JASA* after an excruciatingly long review. Bill empathized but told me that reviews from the *Annals* when he was editor took longer! He also advised on places to submit the article next, and after it had been rejected by several other top journals, each suggesting that it was more appropriate for *JASA*, Bill explained to me that it was acceptable to write to the editor of *JASA* to get his original rejection reconsidered. At the time, I would never have dreamed that I was allowed to be so bold, but given that Bill had been editor of *The Annals of Mathematical Statistics*, I followed his advice and the paper ultimately appeared, after some revision (Fienberg and Gilbert, 1970).

Bill read *The New York Times* daily, and soon I did too, but he clearly read it more carefully than I, often sharing with me clippings or copies of clippings on topics we had discussed that I missed in my morning pass through. I had barely been at Chicago for a few months, but Bill had a mental model for which topics I was working on or in which he thought I would be interested.

Bill always seemed to be editing something, an encyclopedia, a committee report or one of my or someone else's manuscripts. He was always gentle in his suggestions but detailed and probing. I don't think I would have agreed to be a statistical editor for the *International Encyclopedia of the Behavioral and Social Sciences* had Bill not been the editor for the previous version of the encyclopedia during the 1960s. As with so many other activities, Bill convinced me of the importance of such scholarship and exposition for the field of statistics, not just for the social sciences. I also followed Bill's lead in other ways, in academic administration, first as a department chair and later as a dean, and with various professional organizations, such as the IMS of which Bill was president.

Bill's commitment to statistics at the national level left a deep impression on me. His activities took on a new and expanded dimension when he became a member of the President's Commission on Federal Statistics not long after my arrival at Chicago. Moreover, he took the Commission's recommendations to heart and helped found the Committee on National Statistics (CNSTAT) at the National Research Council in the early 1970s. I later joined CNSTAT, but after Bill had rotated off, and in 1980 I became chair, again with Bill's encouragement. One of the most important topics that came up before CNSTAT was methodology for conducting the U.S. decennial census. Bill served on such a committee prior to the 1970 census (Advisory Committee on Problems of Census Enumeration, 1972), even before the creation of CNSTAT. Then he helped to establish a panel on the topic leading up to the 1980 census and I likewise did so for the 1990 census (Citro and Cohen, 1985).

The vexing recurring topic for the census was how to deal with the differential undercount, that is, the difference in the rate of net undercount for Blacks and Whites (later we also included Hispanics and other minority groups). It was here that Bill and I parted ways. I supported the idea of adjustment of census counts for those who were missed—both the development of the methodology for doing this and its actual use as part of the census—and Bill adopted what was for me a somewhat curious stance against adjustment. I say curious because in other contexts he was always pushing for better statistical methodology, documentation of and attention to nonresponse and response errors, and a more prominent role for statistical ideas, whether or not a problem appeared to be statistical. But in the early 1980s, Bill described the decennial census as "a national celebration" (Kruskal, 1984), and he began to articulate a principled position that the census' ceremonial value would be undercut by an adjustment process that was less than perfect. That the census itself was far less than perfect seemed not to matter to him. In fact he would get upset when I'd point out following the 1990 census that almost 10% of all residents were either erroneously omitted or included with error, for example, in the wrong place or double counted.

The 1980s CNSTAT census panel that I participated in strongly supported the adjustment methodology and testing pursued by the Census Bureau staff (Citro and Cohen, 1985). This was viewed as controversial by some and, in 1987, officials in the Department of the Commerce appointed as part of the Republican administration vetoed the use of adjustment for the 1990 census and cancelled the large scale survey on which it was to be based. Several senior statisticians within the Bureau resigned over this decision and a lawsuit followed, brought by states and cities who believed that they would gain from an adjustment. Just before the case was scheduled to go to trial in 1989, the government and the plaintiffs reached a settlement that included a "de novo" decision on adjustment after the census was taken and the creation of an eight-person advisory board to the Secretary of Commerce who would ultimately make the decision on whether or not to adjust the 1990 census results. Four of the eight members were nominated by the Democrats and the other four by the Republicans. Bill was in the latter group and rarely engaged in discussions with the four appointees nominated by the Democrats. Sandy Zabell's (1994) *Statistical Science* interview includes Bill's description of the process, including the fact that his individual report "was brief and non-technical."

Bill really feared that adjustment could introduce errors of its own of unknown magnitude and he valued the census for its iconic national value and feared that an attempt to "correct it" (my phrase and one to which he objected) that did not work and did not command an overwhelming majority of professional support could be a disaster for statistics. What Bill failed to acknowledge at the time and subsequently was the political nature of the entire process and the possibility that his conservative professional stance was being used by others for political purposes. Ultimately, the Secretary of Commerce announced his decision not to adjust, reversing the recommendation to do so from the Census Bureau. The lawsuit resumed and at trial I testified in support of the Bureau's recommendation to adjust.

My relationship with Bill throughout this period was cordial but he seemed unwilling to engage with me in a discussion of the technical details or the empirical evidence for and against adjustment. I couldn't tell if his opposition to adjustment came from a deep-rooted antipathy toward Bayesian methods in which some of the adjustment arguments were couched, from technical arguments raised by others or simply his growing conservatism regarding innovative methodology in such a traditional context since his public statements, including his recommendation to the Secretary of Commerce, were all non-technical and framed in terms of the complexity of census taking. Ultimately, we simply tacitly agreed to disagree on this large statistical issue.

As a junior faculty member at Chicago I changed offices a couple of times, but Bill's office was always close by and I often found myself seated beside him at his desk surrounded by huge stacks of paper discussing some technical issue or seeking advice. I marveled at his ability to retrieve documents and technical papers, almost in mid-conversation, or to share with me additional ones a day or so later. In fact, Bill never stopped sending me copies of letters, clippings, manuscripts or other documents he came across relating to the various topics he thought I was interested in or had worked on years ago. Just today as I worked on this recollection, I also took time to do some office housekeeping and to review material in several stacks that resembled those in Bill's office some 35 years ago. And lo and behold, there were two items from Bill Kruskal, sent I'm not sure when—a reprint of a paper titled "A Question of Religion" on (failed) efforts during the 1950s to collect religious affiliation as part of the census process, and a Xerox of a clipping from *The New York Times*.

Some knew of Bill Kruskal from the Kruskal–Wallis test or the Goodman–Kruskal measures of association, some from his academic and professional leadership. I knew these as well, but I also knew Bill as a scholar with an insatiable appetite for detail and perfection, and more importantly as a mentor and a friend.